\documentclass[10pt,letterpaper,conference]{ieeeconf}

\IEEEoverridecommandlockouts
\usepackage{graphicx}
\usepackage{color}
\usepackage[cmex10]{amsmath}
\usepackage{amssymb}
\usepackage[tight,footnotesize]{subfigure}
\usepackage{float}
\usepackage{multirow}
\usepackage{booktabs}
\usepackage{subeqnarray}
\usepackage{enumerate}
\usepackage{epstopdf}
\usepackage{wrapfig}
\usepackage{soul}
\usepackage[table]{xcolor}
\usepackage{array}
\usepackage[ruled,linesnumbered]{algorithm2e}

\usepackage{cite}
\usepackage[bookmarks=true,colorlinks=true,linkcolor=black,backref=true,
pdfborder={0 0 0},citecolor=blue,urlcolor=blue]{hyperref}

\makeatletter
\let\proof\@undefined			
\let\endproof\@undefined		
\makeatother
\usepackage{amsthm}

\newtheorem{prob}{Problem}

\theoremstyle{definition}

\newcommand{\real}[1][]{\mathbb{R}^{#1}}                                
\newcommand{\nat}[1][]{\mathbb{N}^{#1}}                                 

\newcommand{\defeq}{:=}                                                 
\newcommand{\msub}[1]{_\mathrm{#1}}                                     
\newcommand{\msup}[1]{^\mathrm{#1}}                                     

\newcommand{\eye}[1]{\boldsymbol{I}_{(#1)}}                                          

\newcommand{\df}{\mathrm{d}}                                            

\renewcommand{\geq}{\geqslant}                                          

\newcommand{\intset}[1]{[#1]}

\newcommand{\E}[2]{\mathbb{E}_{#2}\left[ #1 \right]}

\newcommand{\mydef}[1]{{\textit{#1}}}

\newcommand{\ppl}{ath-planning}

\newcommand{\eqnnt}[1]{\hyperref[#1]{(\ref*{#1})}}
\newcommand{\eqnsnt}[2]{\hyperref[#1]{(\ref*{#1})}
	and~\hyperref[#2]{(\ref*{#2})}}
\newcommand{\eqnsernt}[2]{\hyperref[#1]{(\ref*{#1})}--\hyperref[#2]{(\ref*{#2})}}

\newcommand{\eqn}[1]{\hyperref[#1]{Eqn.~(\ref*{#1})}}
\newcommand{\eqns}[2]{\hyperref[#1]{Eqns.~(\ref*{#1})} and~\hyperref[#2]{(\ref*{#2})}}
\newcommand{\eqnser}[2]{\hyperref[#1]{Eqns.~(\ref*{#1})}--\hyperref[#2]{(\ref*{#2})}}
\newcommand{\eqnf}[1]{\hyperref[#1]{Equation~(\ref*{#1})}}
\newcommand{\eqnfs}[2]{\hyperref[#1]{Equations~(\ref*{#1})} and~\hyperref[#2]{(\ref*{#2})}}

\newcommand{\scn}[1]{\hyperref[#1]{Sec.~\ref*{#1}}}
\newcommand{\scns}[2]{\hyperref[#1]{Secs.~\ref*{#1}} and~\hyperref[#2]{\ref*{#2}}}
\newcommand{\scnser}[2]{\hyperref[#1]{Secs~.\ref*{#1}}--\hyperref[#2]{\ref*{#2}}}

\newcommand{\fig}[1]{\hyperref[#1]{Fig.~\ref*{#1}}}
\newcommand{\figs}[2]{\hyperref[#1]{Figs.~\ref*{#1}} and~\hyperref[#2]{\ref*{#2}}}
\newcommand{\figser}[2]{\hyperref[#1]{Figs.~\ref*{#1}}--\hyperref[#2]{\ref*{#2}}}
\newcommand{\figf}[1]{\hyperref[#1]{Figure~\ref*{#1}}}
\newcommand{\figfs}[2]{\hyperref[#1]{Figures~\ref*{#1}} and~\hyperref[#2]{\ref*{#2}}}
\newcommand{\figfser}[2]{\hyperref[#1]{Figures~\ref*{#1}}--\hyperref[#2]{\ref*{#2}}}

\newcommand{\tbl}[1]{\hyperref[#1]{Table~\ref*{#1}}}
\newcommand{\tbls}[2]{\hyperref[#1]{Tables~\ref*{#1}} and~\hyperref[#2]{\ref*{#2}}}
\newcommand{\tblser}[2]{\hyperref[#1]{Tables~\ref*{#1}}--\hyperref[#2]{\ref*{#2}}}

\newcommand{\apx}[1]{\hyperref[#1]{Appendix~\ref*{#1}}}

\newcommand{\prb}[1]{\hyperref[#1]{Problem~\ref*{#1}}}
\newcommand{\prp}[1]{\hyperref[#1]{Prop.~\ref*{#1}}}
\newcommand{\prpf}[1]{\hyperref[#1]{Proposition~\ref*{#1}}}

\newcommand{\algoref}[1]{\hyperref[#1]{Algorithm~\ref*{#1}}}

\newcommand{\thmref}[1]{\hyperref[#1]{Theorem~\ref*{#1}}}
\newcommand{\thmsref}[2]{\hyperref[#1]{Theorems~\ref*{#1}} and~\hyperref[#2]{\ref*{#2}}}
\newcommand{\thmserref}[2]{\hyperref[#1]{Theorems~\ref*{#1}}--\hyperref[#2]{\ref*{#2}}}

\newcommand{\algline}[1]{\hyperref[#1]{Line~\ref*{#1}}}
\newcommand{\alglines}[2]{\hyperref[#1]{Lines~\ref*{#1}} and~\hyperref[#2]{\ref*{#2}}}
\newcommand{\alglineser}[2]{\hyperref[#1]{Lines~\ref*{#1}}--\hyperref[#2]{\ref*{#2}}}

\renewcommand{\vec}[1]{\boldsymbol{#1}}

\newcommand{\graph}{\mathcal{G}}
\newcommand{\verts}{V}
\newcommand{\edges}{E}

\def\wsp{\mathcal{W}}
\def\threat{c}

\def\vertInit{i\msub{s}}
\def\vertGoal{i\msub{g}}

\newcommand{\xState}{\vec{x}}
\newcommand{\qConfig}{\vec{q}}
\newcommand{\zMeas}{\vec{z}}
\newcommand{\QProcCovar}{Q}
\newcommand{\PEECovar}{P}
\newcommand{\RMeasCovar}{R}
\newcommand{\gridPath}{\vec{\pi}}
\newcommand{\speedactor}{u\msub{ego}}
\newcommand{\speedsensor}{u\msub{sen}}
\newcommand{\nSensor}{N\msub{s}}
\newcommand{\nGridPt}{N\msub{g}}
\newcommand{\paramVec}{\boldsymbol{\Theta}}
\newcommand{\HMeas}{H}

\newcommand{\SensorCost}{f}

\newcommand{\egoPosition}[1]{\xState\msub{ego}^{#1}}
\def\constA {\alpha}
\def\constB{\gamma}

\def\SensorConfigReward{r}
\def\integralCost{\mathcal{J}}

\title{\Large\bf Actively Coupled Sensor Configuration and Planning \\ in Unknown Dynamic Environments}
\author{Prakash Poudel$^{\ast}$, Jeffrey DesRoches$^{\ast}$, and Raghvendra V. Cowlagi$^{\dagger}$   
\thanks{$^{\ast}$Graduate Research Assistant, Aerospace Engineering.}%
\thanks{$^{\dagger}$Associate Professor, Aerospace Engineering, Worcester Polytechnic Institute, Worcester, MA, USA. Corresponding author.}%
}

\begin{document}
\maketitle
\begin{abstract}
	We address the problem of path-planning for an autonomous mobile vehicle, 
	called the ego vehicle, in an unknown and time-varying environment. The objective
	is for the ego vehicle to minimize exposure to a spatiotemporally-varying unknown
	scalar field called the threat field. Noisy measurements of the threat field are 
	provided by a network of mobile sensors. We address the problem of optimally
	configuring (placing) these sensors in the environment. To this end, we propose 
	sensor reconfiguration by maximizing a reward function composed of three different
	elements. First, the reward includes an information measure that we call 
	context-relevant mutual information (CRMI). Unlike typical sensor placement
	techniques that maximize mutual information of the measurements and environment state,
	CRMI directly quantifies uncertainty reduction in the ego path cost while it moves
	in the environment. Therefore, the CRMI introduces active coupling between
	the ego vehicle and the sensor network. Second, the reward includes a
	penalty on the distances traveled by the sensors. Third, the reward includes
	a measure of proximity of the sensors to the ego vehicle. Although we do
	not consider communication issues in this paper, such proximity is of relevance
	for future work that addresses communications between the sensors and the ego vehicle.
	We illustrate and analyze the proposed technique via numerical simulations.	
\end{abstract}

\section{Introduction}

Many envisioned applications of autonomous mobile agents, such as emergency first response 
after natural disasters, involve operations in dynamic and uncertain environments. 
In such applications, an autonomous agent
may need to navigate through adverse environmental conditions, to which we would like 
to minimize exposure. We refer to these adverse conditions in aggregate as a \mydef{threat field}. 
The threat field is a spatiotemorally-varying scalar field that represents unfavorable conditions 
such as extreme weather, harmful chemical substances, or radiation. The agent may need 
to traverse the environment while balancing the competing objectives of reducing 
exposure to the threat and reducing mission completion time. In the rest of this
paper, we refer to this agent as the \mydef{ego vehicle.}

The ego vehicle is supported by a spatially distributed network of mobile sensors. 
These sensors collect real-time data of the threat field, which may be used by the ego vehicle.
In applications such as emergency first response, the availability of mobile sensors, such as 
unmanned aerial vehicles (UAVs), to gather information over large areas may be limited.
With this motivation, we focus on the problem of p\ppl\ for the ego vehicle using a minimal
number of sensor measurements.

This problem is naturally related to several different disciplines in the literature
including p\ppl\ under uncertainty and state estimation. Utilizing sensor data from a
mobile sensor network also introduces the problem of optimal sensor placement, which 
is necessary to ensure that ego vehicle has adequate information about the threat field 
with as few measurements as possible.

Classical approaches to path-planning include cell decomposition, probabilistic roadmaps, 
and artificial potential field techniques \cite{LaValle2006}, \cite{Patle2019}. 
Dijkstra's algorithm, $\textrm{A}^{\ast}$, and their variants are branch-and-bound 
optimization methods that use heuristics to systematically explore the search space and 
identify the shortest path. While classical path-planning methods are powerful, they are 
inherently limited by the accuracy of the environment's available information. 
An accurate representation of the environment is difficult if the environment's states 
or dynamics are unknown. Learning-based approaches, particularly deep reinforcement learning 
(RL)~\cite{wen2024drl, qin2023deep} are gaining attention for their ability to handle complex 
and uncertain environments.

Sensor data are always noisy and often incomplete, and therefore it is necessary to apply
probabilistic estimation techniques to obtain accurate information. The literature on 
estimation includes various Bayesian techniques, such as the Kalman filter~\cite{Lewis2017optimal}, 
maximum likelihood estimator~\cite{myung2003tutorial}, and Bayesian filter~\cite{thrun2006}. 
The use of the extended Kalman filter (EKF), the unscented Kalman filter (UKF)~\cite{Julier2004}, 
or the particle filter~\cite{deng2021poserbpf} is prevalent for nonlinear dynamical systems.

Various sensor placement strategies have been employed depending on the type of 
application and parameters that need to be measured. Greedy approaches that utilize 
information-based metrics have been explored in the 
literature~\cite{kohara2020sensor,Soderlund2019}. Machine learning techniques for 
sensor placement aim to achieve efficient sensing with the fewest possible sensors and 
measurements~\cite{nayak2021routing,Wang2020,hoffmann2020}. Information-theoretic 
sensor placement methods employ performance metrics such as the Fisher information matrix 
(FIM)~\cite{Kangsheng2006}, entropy~\cite{Wang2004}, Kullback-Leibler (KL) 
divergence~\cite{liu2024network}, mutual information~\cite{Krause2008,Adurthi2020}
to maximize the amount of useful information collected from the environment.
Although it is commonly studied in the mobile sensor network literature,
the problem of accounting for sensor reconfiguration costs is relatively less
studied for sensor placement. Reconfiguration cost becomes important when multiple
iterations of the sensor configuration are implemented in a time-marching environment.
Some prior research includes the reconfiguration cost of the sensor network
topology~\cite{leong2014network}, or the total energy consumption of the sensor 
network~\cite{ramachandran2015measuring,grichi2017new}.

In this paper we consider the problem of optimal sensor configuration coupled with 
p\ppl\ in an unknown dynamic environment. More specifically, the objective is to strategically 
place sensors in locations that provide information of the most relevance to the 
p\ppl\ problem. For p\ppl\ under uncertainty, the most relevant information is that 
which can maximally reduce uncertainty about the path cost. This is a relatively new
research problem in that it departs from the usual separation of estimation and 
planning/control and instead explicitly couples sensor configuration to planning.

Earlier studies have addressed this problem in the context of static environments. 
A heuristic task-driven sensor placement approach called the interactive planning and 
sensing (IPAS) for static environments is reported in~\cite{Cooper2019}. 
This approach is reported to outperform several decoupled sensor placement 
strategies by reducing the number of configurations necessary to achieve 
near-optimal paths. Sensor configuration for both location and 
field-of-view has also been investigated for static fields~\cite{Laurent2023}. 
Optimal sensor placement in a time-varying threat field that is  based on optimizing 
a novel path dependent information measure, referred to as \mydef{context-relevant mutual information}
(CRMI) is presented in our previous work~\cite{poudel2024coupled}. 
A modified sensor placement method that maximizes a weighted sum of 
CRMI and a reward for reducing distances traveled by sensors
is reported in~\cite{Poudel-Cowlagi-Scitech2025}.

All of this prior research~\cite{poudel2024coupled, Poudel-Cowlagi-Scitech2025} 
is based on an iterative process that identifies the optimal sensor configuration, 
obtains the sensor measurements, updates the threat estimates, and plans path accordingly. 
This iterative process terminates when the path cost variance falls below a 
user-specified threshold. \emph{Crucially, the ego vehicle is assumed to move only after the 
optimal path has been identified.} In other words, ego vehicle is assumed to ``wait'' at its initial
location until this iterative process converges. Whereas ths assumption is relevant 
for the preliminary development of the coupled sensor configuration and planning 
methods, it is not acceptable from a practical perspective in time-varying environments.
Waiting at the initial location can make the planning outcome obsolete due to changes in 
the environment.

The novelty of this work is that we consider a situation where the ego vehicle moves
simultaneously with the sensor configuration process. 
Therefore, we refer to the proposed technique as \emph{actively coupled sensor 
configuration and p\ppl} (A-CSCP). We consider the optimization of a new 
sensor configuration objective function that depends not only on CRMI and the distance
traveled by sensors, but also on the relative distances between the sensors and the
ego vehicle as it moves.

The rest of the paper is organized as follows. 
In~\scn{sec-problem}, we introduce the elements of the problem formulation.
In~\scn{sec-cscp}, we present the new objective function and the A-CSCP iterative
algorithm. In~\scn{sec-results}, we present illustrative examples and
comparative results, and conclude the paper in~\scn{sec-conclusions} with
comments on future work.

\section{Problem Formulation}
\label{sec-problem}
Let $\real$ be the set of real numbers, $\nat$ the set of natural numbers, and
$\eye{N}$ the identity matrix of size $N$.
For any $N\in\nat$, let $\intset{N}$ denote \{1, 2, \ldots, $N$\}.

In a closed square region, $\wsp\subset\real[2]$, referred to as the workspace, 
a mobile agent operates alongside a network of spatially distributed sensors. 
The workspace is divided into a grid of $\nGridPt$ uniformly spaced points, with each point 
assigned coordinates $\xState_{i}$ in a prespecified Cartesian coordinate axis system, for each 
$i\in{\nGridPt}$. The distance between adjacent grid points is denoted by $\delta$. 
The agent navigates this grid according to the ``4-way adjacency rule", which means movement 
is restricted to adjacent points in the upward, downward, leftward, and rightward directions. 
This navigation problem is framed as a graph search over a graph $\graph = (\verts,\edges)$, 
where $\verts = \intset{\nGridPt}$ is the set of vertices, and $\edges$ 
is the set of edges connecting geometrically adjacent vertices. 
Each vertex in $\verts$ is uniquely associated with a grid point.

A \textit{threat field}, denoted as $\threat:\wsp\times\real_{\geq0}\rightarrow\real_{>0}$, 
is a time-varying scalar field that takes strictly positive values, indicating regions with 
higher intensity that are potentially hazardous and unfavorable. The ego vehicle is required 
to move from a start vertex $\vertInit\in \verts$ to a goal vertex $\vertGoal\in \verts$, 
following a path $\gridPath = \{i_{0},i_{1},\ldots,i_{L}\}$, where $i_{0} = \vertInit$ 
and $i_{L} = \vertGoal$ for some $L \geq 1.$ 
The ego vehicle is assumed to move at a constant speed, $\speedactor$. 
Each transition between vertices incurs 
a cost that is determined by the threat field exposure. 
The cost associated with traversing adjacent 
grid points is denoted by a scalar function 
$g:\edges \times \real_{\geq0} \rightarrow\real_{>0}$ defined as
\begin{equation}
	g((i,j),t) = \threat(\xState_{j},t) \mbox{\qquad for } (i,j)\in \edges.
\end{equation} 
The overall threat exposure $J(\gridPath)$ along the path is the sum of 
the edge transition costs, i.e., $J(\gridPath) \defeq \delta
\sum_{\ell=1}^{L} g((i_{\ell-1},i_{\ell}),\ell\Delta t_{s})$. 
The main goal is to identify a path $\gridPath^{*}$ with 
a minimum cost.

Because the threat field is unknown, it is necessary to estimate its values. 
To this end, a mobile sensor network of $\nSensor$ sensors is deployed to measure the 
intensity of the threat field at various points. We assume that $\nSensor \ll \nGridPt,$
i.e., that we have a relatively small number of sensors available. 
Each sensor in the network is assumed to move at a constant speed 
$\speedsensor > \speedactor.$ 
The sensor measurements are denoted $\zMeas(t;\qConfig) = 
\{z_{1}(t;\qConfig),z_{2}(t;\qConfig),\ldots,z_{N_{s}}(t;\qConfig)\}.$
The sensors are positioned at specific grid points, and the set of 
grid points where sensors are located is called the \textit{sensor configuration}, 
$\qConfig = \{q_{1},q_{2},\ldots,q_{N_{s}}\}\subset[\nGridPt]$.

The threat field is represented by a parametric model of the form
$\threat(\xState,t) \defeq 1 + \sum_{n=1}^{N_{P}}\theta_{n}(t)\phi_{n}(\xState) = 1 + 
\boldsymbol\Phi^{\intercal}(\xState)\paramVec(t)$, where $\boldsymbol\Phi(\xState)$ 
is a vector of spatial basis functions, defined as 
$\boldsymbol\Phi(\xState):=[\phi_{1}(\xState)\ldots\phi_{{N_{P}}}(\xState)]^{\intercal}$. 
Here, $N_{P}$ represents the number of parameters used to model the threat field. 
For each $n\in[N_{P}]$, $\phi_{n}(\xState)$ is modeled as a Gaussian function,
$\phi_{n}(\xState) \defeq 
\exp(-(\xState-\overline{\xState}_{n})^{\intercal}(\xState-\overline{\xState}_{n})/2a_{n})$. 
The constants 
$a_{n}\in\real_{>0}$ and $\overline{\xState}_{n}\in \wsp$ are prespecified, 
ensuring that the combined regions 
of influence of the basis functions sufficiently 
cover the entire workspace~$\wsp.$ 
The time-varying parameter 
$\paramVec(t)\defeq[\theta_{1}(t)\ldots\theta_{{N_{P}}}(t)]^{\intercal}$ 
is unknown and is to be estimated using the sensor measurements $\zMeas.$

The temporal evolution of the threat in discrete form
is defined using the linear system dynamic model,   
\begin{align}
	\paramVec_{k} &= A \paramVec_{k-1} + \boldsymbol\omega_{k-1},
	\label{eq-threat-evolution}
	\end{align}
\noindent
where the matrix $A$ represents some known evolution of threat parameters
and $\boldsymbol\omega_{k} \sim \mathcal{N}(0, \QProcCovar)$, 
with $\QProcCovar \defeq \sigma_{P}\eye{N_{P}}$
for each $k \in \nat.$ 
In this paper we assume that the system~\eqnnt{eq-threat-evolution}
is stable.

The measurements obtained from each sensor are
modeled by $\zMeas_{k} \defeq c(\xState_{{q}_{k}},t) + \boldsymbol\eta_{k} = 
\HMeas_{k}(\qConfig)\paramVec_{k} + \boldsymbol\eta_{k}$, where the matrix $$\HMeas_{k}(\qConfig) = 
\left[\boldsymbol\Phi(\xState_{\qConfig_{k,1}}) \qquad \boldsymbol\Phi(\xState_{\qConfig_{k,2}}) \ldots 
\boldsymbol\Phi(\xState_{\qConfig_{k,\nSensor}})\right]^{\intercal},$$ and $\boldsymbol\eta_{k} \sim 
\mathcal{N}(0, \RMeasCovar)$ is zero mean measurement noise with covariance $\RMeasCovar \succ 0$.
Using these measurements and a Bayesian estimation algorithm such
as the Kalman filter, we can find stochastic estimates of the threat parameter
with mean value $\widehat{\paramVec}(t)$ and estimation error covariance $\PEECovar.$

For any path, $\gridPath = 
\{i_{0},i_{1},\ldots,i_{L}\}$ in $\graph$, the cost of the path is 
\begin{align*}
	J(\gridPath) \defeq L + 
	\delta\textstyle{\sum_{\ell=1}^{L}} \boldsymbol\Phi^{\intercal}(\xState_{\ell})\paramVec(t).
\end{align*}
The cost $J$ becomes a random variable with distribution dependent on the
estimate $\paramVec.$ The problem of interest in this paper is then defined as follows.

\begin{prob}
	For a prespecified ego vehicle start and goal vertices
	$\vertInit,\vertGoal \in \verts,$ with the ego vehicle 
	and finite sensors moving at constant speeds $\speedactor$ 
	and $\speedsensor$ respectively, find sensor configurations 
	$\qConfig^*$ and a path $\gridPath^{*}$ with minimum expected
	threat exposure $\E{J(\gridPath^{*})}{}.$
	\label{main-prob}
\end{prob}

Because the number of sensors $\nSensor$ is small relative to the 
number of grid points $\nGridPt,$ we seek an iterative solution 
to~\prb{main-prob}, whereby the sensor configuration and planned
path are iteratively updated. These iterations occur while the 
ego vehicle moves, and therefore the iterative computations must 
incorporate the ego vehicle's changing location.

\section{Active Coupled Sensing and Planning}
\label{sec-cscp}

Active coupled sensor configuration and path-planning (A-CSCP) 
is the proposed approach to solve Problem 1 in a time-varying environment. 
At each time step, multiple sensors and an ego vehicle move at constant speeds 
towards their objectives. The sensors gather pointwise measurements of the threat field. 
The optimal sensor configuration is determined based on an information 
measure called \mydef{context-relevant mutual information} (CRMI), which we
describe in further detail below. As each sensor reaches its next configuration, 
it takes a measurement and reports to a central server, which in turn 
updates the threat field estimate. 
Then, the optimal path for the ego vehicle is updated and a new configuration 
for the sensor is determined.

Any estimator can be employed to estimate the state parameters. If 
the threat parameter model is linear, as we assume in \eqnnt{eq-threat-evolution},
then we may use a simple Kalman filter. For generality and applicability to
future work with nonlinear evolution models, we may utilize an 
Unscented Kalman Filter (UKF~\cite{Julier2004}).

As the ego vehicle travels and arrives at each grid point along its planned path, 
it replans its path based on the latest threat estimate. 
This coordinated interaction between sensor reconfiguration and ego vehicle movement
provides continuous adaptation to the evolving threat environment.
This process continues until the ego vehicle reaches the goal vertex.
In what follows, we provide details of this iterative process, analysis, and an illustrative example.

\subsection{Context Relevant Mutual Information}
We define the \mydef{context-relevant mutual information} (CRMI) to quantify 
the information shared between the path cost and sensor measurements. CRMI
is the novel and crucial coupling between sensor configuration and p\ppl.
The CRMI becomes maximum at the spatial locations of most relevance to p\ppl, 
disregarding areas that are far from the intended path. In other words, placing
sensors at maximum CRMI locations is likely to reduce the path cost uncertainty
by the largeest amount.

For any path $\gridPath$, the expected cost is $\widehat{J}(\gridPath) := L + 
\delta\sum_{\ell=1}^{L} \boldsymbol\Phi(\xState_{\ell})^{\intercal}\widehat{\paramVec}(t).$
Because the estimate $\widehat{\paramVec}, \PEECovar$ is Gaussian and because the 
path cost is a linear function of the parameter estimate, the path cost r.v. is also
Gaussian. Therefore, the joint PDF $p(J_{k},\zMeas_{k})$ of the path cost and measurement~is
\begin{align*}
	p(J_{k},\zMeas_{k}) &= \mathcal{N}\left(\begin{bmatrix} J_{k}\\ \zMeas_{k} \end{bmatrix} : 
	\begin{bmatrix} \widehat{J}_{k|k-1}\\ \widehat{\zMeas}_{k} \end{bmatrix}
	, \begin{bmatrix} {\PEECovar}_{J J_{k|k-1}} & \PEECovar_{{J \zMeas}_{k|k-1}} \\ 
	\PEECovar^{\intercal}_{{J \zMeas}_{k|k-1}} & \PEECovar_{{\zMeas\zMeas}_{k|k-1}} \end{bmatrix}\right).
\end{align*}
The variance of the path cost is
\begin{align}
	{\PEECovar}_{J J_{k|k-1}} &\defeq
	\mathbb{E}\left[\left(J(\gridPath) - \widehat{J}(\gridPath) \right)^{2}\right] \nonumber \\
	&= 
	\mathbb{E}\left[\left(\delta\sum_{\ell=1}^{L} 
	\boldsymbol\Phi^\intercal(\xState_{\ell}) \left(\paramVec(t) - 
	\widehat{\paramVec}(t)\right) \right)^{2}\right], \nonumber \\
    \notag
    &= 
    \delta^{2} \sum_{\ell=1}^{L}\left(\boldsymbol\Phi(\xState_{\ell})^{\intercal}
    \PEECovar_{k_{\ell}}\boldsymbol\Phi(\xState_{\ell}) \right) \\
    & 	\hspace{-3ex}  + 2\delta^2 \sum_{\ell < m, 
    ~~\ell,m\in[L]}^{L}\left(\boldsymbol\Phi(\xState_{\ell})^{\intercal}
	\PEECovar_{k_{\ell m}}\boldsymbol\Phi(\xState_{m})
     \right). \label{eq-covar1}
\end{align} 

To compute $\PEECovar_{J J_{k|k-1}}$, we first need to determine
$\boldsymbol\Phi$  and the error covariance $\PEECovar$ for each
grid point $\gridPath_{l}$ along the path. $\PEECovar_{k_{l}}$ and
$\PEECovar_{k_{lm}}$ are determined by propagating the UKF prediction
steps for a time steps for traversing between grid points.
The covariance of the measurement and the cross covariance between
the path cost and the measurement random vector are then formulated as
\begin{align}
	\notag
	\PEECovar_{{J \zMeas}_{k|k-1}} &\defeq \mathbb{E}\left[\left(\zMeas 
	-\widehat{\zMeas}\right)\left(J(\gridPath) - \widehat{J}(\gridPath)\right)\right]\\ 
	&= 
	\delta\sum_{l=1}^{L}\left(\boldsymbol\Phi(\xState_{\gridPath_{l}})^{\intercal}\PEECovar_{k_{l}} 
	\right)\HMeas_{k}^{\intercal}(\qConfig), \\
	%
	\PEECovar_{{\zMeas\zMeas}_{k|k-1}} &= 
	\HMeas_{k}(\qConfig)\PEECovar_{\paramVec\paramVec_{k|k-1}}\HMeas_{k}^{\intercal}(\qConfig) + 
	\RMeasCovar_{k}. \label{eq-covar3}
\end{align}
 Here, $\PEECovar_{\paramVec\paramVec_{k|k-1}}$ is the priori state covariance that is obtained from the UKF algorithm. Finally, the CRMI is calculated as
 \begin{align}
 \hspace{-2ex}I({J}_{k};\zMeas_{k}(\qConfig)) &= \nonumber \\
 & \hspace{-7ex}
 \frac{1}{2}\log\left(\frac{|\PEECovar_{JJ_{k|k-1}}|}{|\PEECovar_{JJ_{k|k-1}} - 
 \PEECovar_{{J \zMeas}_{k|k-1}} \PEECovar^{-1}_{{\zMeas\zMeas}_{k|k-1}} \PEECovar^{\intercal}_{{J 
 \zMeas}_{k|k-1}}|}\right).
	\label{eq-crmi}
\end{align}

\subsection{Sensor Reconfiguration Cost}

In a previous work~\cite{poudel2024coupled}, we showed that CRMI is a submodular
function. Optimization of submodular functions is computationally convenient
because a greedy algorithm is known to converge to a near-optimal solution
with bounded worst-case suboptimality~\cite{nemhauser1978analysis}. In the present context,
greedy optimization implies that we find the optimal configuration of each sensor
one at a time. Based on this observation from previous work, we consider greedy
optimization, only, in the rest of this work.

Sensor reconfiguration cost represents the cost associated with relocating 
mobile sensors within the environment. In this context, we define the sensor 
reconfiguration cost based on two distance components, described as follows.

At the $\ell\msup{th}$ sensor configuration within the 
A-CSCP algorithm (described in the next
subsection) and for the $j\msup{th}$ sensor, the first component, 
denoted $d_1,$ is the Euclidean 
distance between the sensor's current location $q^{\ell*}_j$ and its 
new candidate location $q^{\ell+1}_j \in \intset{\nGridPt} 
\backslash \qConfig^{\ell*},$ i.e., 
\begin{align*}
	d_1(q^{\ell+1}_j) \defeq \| q^{\ell+1}_{j} - q^{\ell*}_j \|.
\end{align*}

The second component $d_2$ is the distance between the sensor's new candidate 
location $q^{\ell+1}_{j}$ and the position $\egoPosition{\ell}$ of the ego vehicle's 
next planned grid point, i.e., 
\begin{align*}
	d_2(q_{j}^{\ell + 1}) = \| q_{j}^{\ell + 1} - \egoPosition{\ell + 1}\|.
\end{align*}
In other words, $\egoPosition{\ell+1} \in \wsp$ is the location associated with
the second vertex in the path planned at iteration $\ell.$
Considering this distance in the sensor reconfiguration cost allows the 
sensor placement to consider the proximity of the sensors to the ego vehicle.

Next, we define the weighted sum
\begin{align*}
	d(q_{j}^{\ell + 1}) \defeq \constB d_1(q_{j}^{\ell + 1}) + (1-\constB) d_2(q_{j}^{\ell + 1}).
\end{align*}
The constant $\constB$ provides a trade-off between the two distance 
components $d_1$ and $d_2$ in the sensor reconfiguration cost. 
When $\constB=1$, the sensor configuration cost is solely based on the 
Euclidean distances traveled by the sensors.
Similarly, when  $\constB=0$, the sensor configuration cost considers 
the distance between sensor’s current 
location and the ego vehicle next planned location, only.

Next, we define the sensor reconfiguration cost as
\begin{align}
	\SensorCost(q_{j}^{\ell + 1}) &\defeq 
	\min_{q \in \intset{\nGridPt} \backslash \qConfig^{\ell*}} 
		\left\{ d(q) \right\} - d(q_{j}^{\ell + 1}).
	\label{eq-reconf-cost}
\end{align}
Finally, we define the reward function
\begin{align}
	\SensorConfigReward(q_{j}^{\ell + 1}) \defeq I({J}; \vec{z}(q_{j}^{\ell + 1})) + 
	\constA \SensorCost(q_{j}^{\ell + 1})
	\label{eq-main-reward}
\end{align}
to be maximized for finding the next configuration for sensor~$j.$
Note that, due to the $\min$ term in \eqnnt{eq-reconf-cost},
$\SensorCost$ is always nonpositive, i.e., $\SensorCost$
reduces the reward associated with CRMI. In other words, $\SensorConfigReward$
is a reward that encodes a balance between the most path-relevant
informativeness, distance traveled by sensors, and proximity between 
sensors and the ego vehicle. The term $I({J}; \vec{z}(q_{j}^{\ell + 1}))$
is to be understood as the CRMI for the sensor configuration obtained 
by moving the $j\msup{th}$ sensor to $q_{j}^{\ell + 1})$ while
keeping all other sensors fixed.

The next sensor configuration is then determined as
\begin{align}
	q_{j}^{(\ell + 1)*} \defeq \max_{q \in \intset{\nGridPt}\backslash \qConfig^{\ell^*}}
	\left\{ \SensorConfigReward(q) \right\}.
	\label{eq-sensorcost}
\end{align}

The constant $\constA$ in \eqnnt{eq-main-reward} is a normalizing factor that 
balances the trade-off between maximizing the CRMI metric and 
minimizing the sensor reconfiguration cost. {To be precise:}
\begin{align}
	\constA \defeq \frac{\max_q \left\{ 
		I({J};\vec{z}(q))\right\}}{\max_q 
		\left\{ d(q) \right\} -	\min_q \left\{ d(q) \right\}  }.
		\label{eq-normalize}
\end{align}
The two maxima and the minimum in \eqnnt{eq-normalize} are calculated over 
the set of feasible configurations ${ \intset{\nGridPt} \backslash \qConfig^{\ell*}}.$

\subsection{Active CSCP Algorithm}

\begin{algorithm}
	\SetAlgoLined
	Set time step $k=0,$ and $\widehat{\paramVec}_{0} = \mathbf{0},$ 
	and $\PEECovar_{0}=\chi\eye{N_{P}}$
	
	
	Initialize sensor configuration $\qConfig^{0*} \subset \intset{\nGridPt}$
	
	Initialize ego vehicle location to at $\vertInit$
	
	Obtain $\zMeas( t_0; \qConfig^{0*})$ and update $\widehat{\paramVec}_{0}, \PEECovar_{0}$
	
	Find $\gridPath^{*}_{0}= \arg\min (\widehat{J}_{0}(\gridPath) ) $
	
	
	For each $j\in\intset{\nSensor},$ find optimal sensor configuration $q^{1*}_j 
	\defeq \arg\max_{q} (\SensorConfigReward(q))$
	
	
	\While {ego vehicle position $\neq(\xState_{\vertGoal})$}{
		
		Move ego vehicle along $\gridPath{}^*$ at speed $\speedactor$
		
		\If {ego vehicle position is at $(\xState_{i_0})$}{
			
			$i_0 = i_1$						
			
		} 
		
		\For{$j = 1:\nSensor$}{
			Move sensor $j$ toward $\xState_{q^{\ell + 1}_j}$ at speed $\speedsensor$
			
			\If {sensor $j$ position is $\xState_{q^{\ell+1}_j}$}{
				
				$\ell=\ell+1$
				
				Obtain $\zMeas_{t_\ell; \qConfig^{\ell*}}$ and 
				update $\widehat{\paramVec}_{k}, \PEECovar_{k}$
				
				Find $\gridPath^{*}_{k}= \arg\min (\widehat{J}_{k}(\gridPath) ) $
				
				
				Find optimal sensor configuration $q^{\ell*}_{j} = 
				\arg\max_{q} (\SensorConfigReward(q))$
				\label{line-config}
				
			} 
			
		}
		{$k = k + 1$}
	}
	\caption{A-CSCP Algorithm}
	\label{alg-A-CSCP}
\end{algorithm}

The active coupled sensing and planning (A-CSCP) algorithm 
described in \algoref{alg-A-CSCP} initializes with the prior 
$\widehat{\paramVec}_{0} = \vec{0}$ and $\PEECovar_{0}=\chi\eye{N_{P}}$,
where $\chi$ is a large arbitrary number. The ego vehicle's start vertex 
is set as $\vertInit$, and the final goal vertex is $\vertGoal$. 
The initial sensor configuration $\qConfig^{0*}$ is
such that the sensors are placed at the immediate neighbors of $\vertInit.$ 
The sensors immediately take measurements, and update the estimates of $\paramVec.$ 
An optimal path $\gridPath^{*}_{0} = \{i_0, i_1, \ldots, i_L\}$ 
for the ego vehicle is determined (with $i_0 = \vertInit$),
and the sensors then calculate their next configuration 
$\qConfig^{1*}$, using the aforesaid greedy method.

The ego vehicle and sensors then move at predefined speeds
$\speedactor$ and $\speedsensor$, respectively toward their 
respective next vertices. 
When the ego vehicle reaches the next vertex, it adopts 
the latest optimal path, which is determined using the most 
recent threat field estimates based on the latest sensor measurements. 
The ego vehicle's starting vertex is updated to the next vertex, 
$i_1$ along the optimal path. Since the CRMI for the portion of 
the path already traversed is irrelevant for future path planning,
the searching algorithm is restricted to considering only the relevant
future paths, meaning that the previously traveled path cannot be altered.

For each sensor, the optimal sensor configuration is determined by 
maximizing the reward $\SensorConfigReward$ defined in \eqnnt{eq-main-reward}. 
When a sensor reaches its next configuration vertex, it takes a new measurement and 
the threat parameter estimate $\paramVec$ is updated. 
Again a optimal path is calculated, and the new configuration for that 
sensor is determined. This process continues until the ego vehicle reaches 
the goal vertex $\vertGoal.$

\section{Results and Discussion}
\label{sec-results}

This section first provides an illustrative example of the proposed A-CSCP
method. Next, we perform comparative analysis based on different schemes of 
sensor placement discussed earlier.
All numerical simulations are performed within a square workspace 
$\wsp = [-1, 1] \times [-1, 1]$ using non-dimensional units.

\subsection{Illustrative Example}

\def\thisfigwidth{0.46\columnwidth}
\begin{figure}
	\centering
	\subfigure[$i=1$]{
		\includegraphics[width=\thisfigwidth]{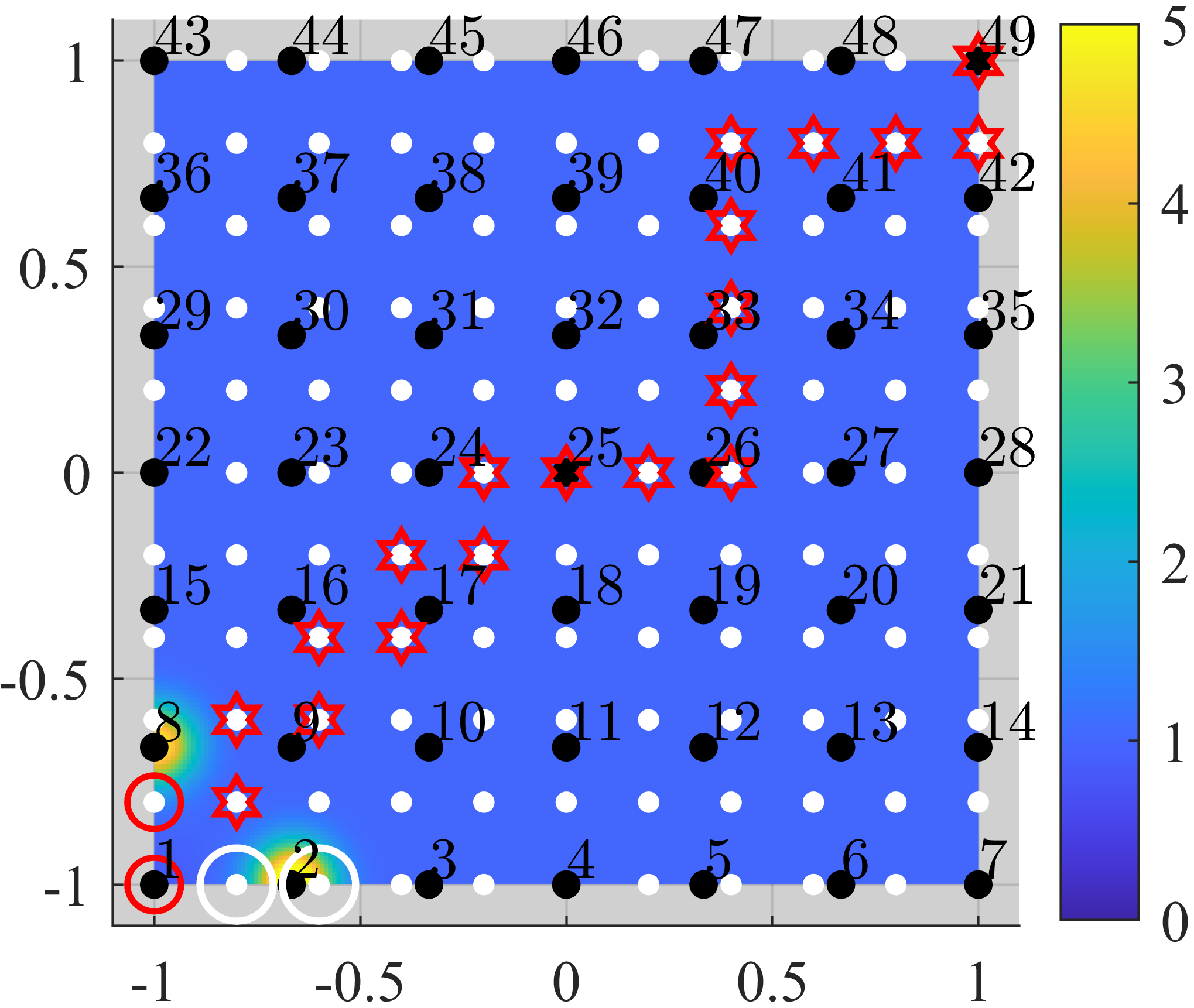}
	}
	\subfigure[$i=6$]{
		\includegraphics[width=\thisfigwidth]{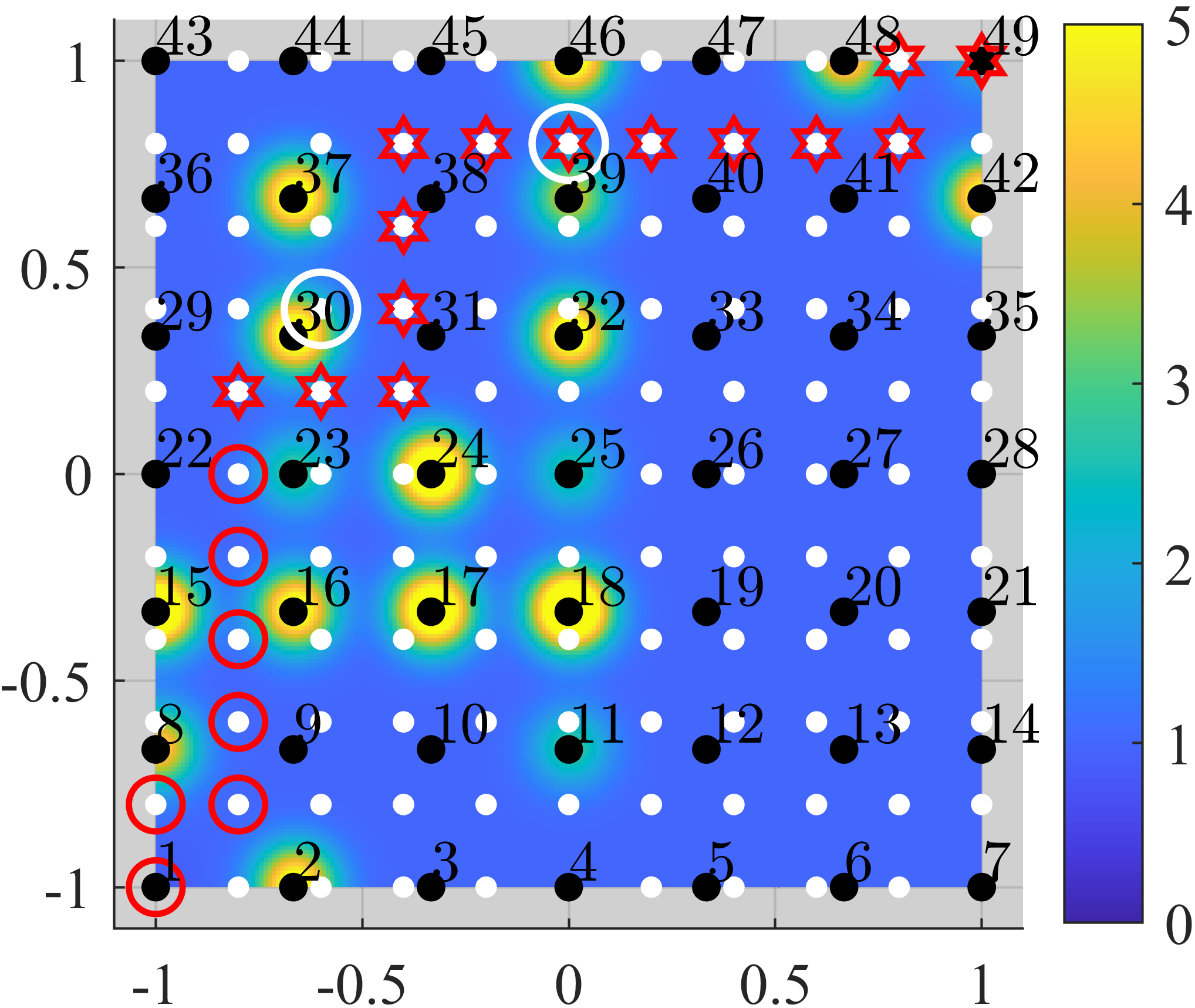}
	}
	\subfigure[$i=13$]{
		\includegraphics[width=\thisfigwidth]{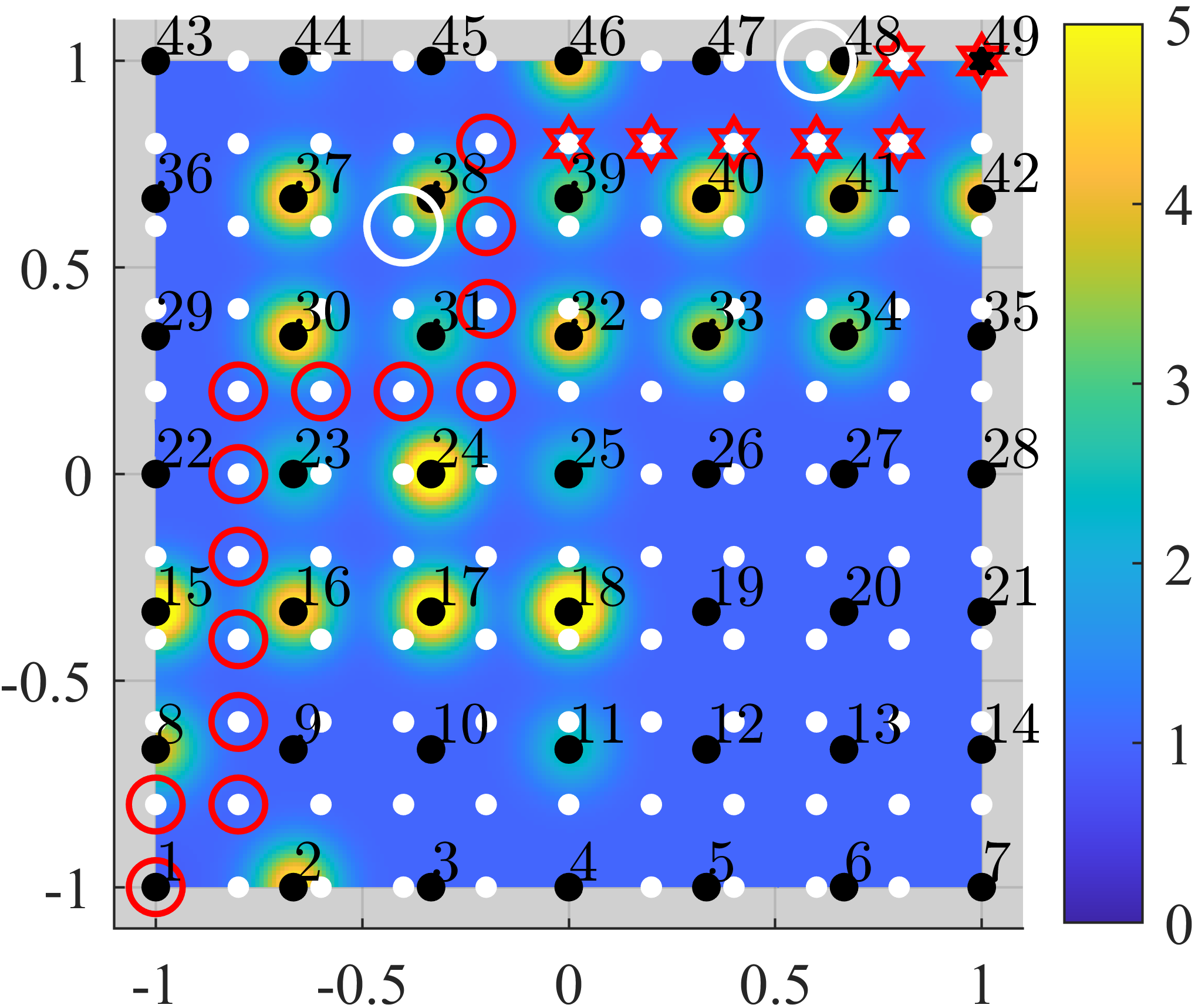}
	}
	\subfigure[$i=19$]{
		\includegraphics[width=\thisfigwidth]{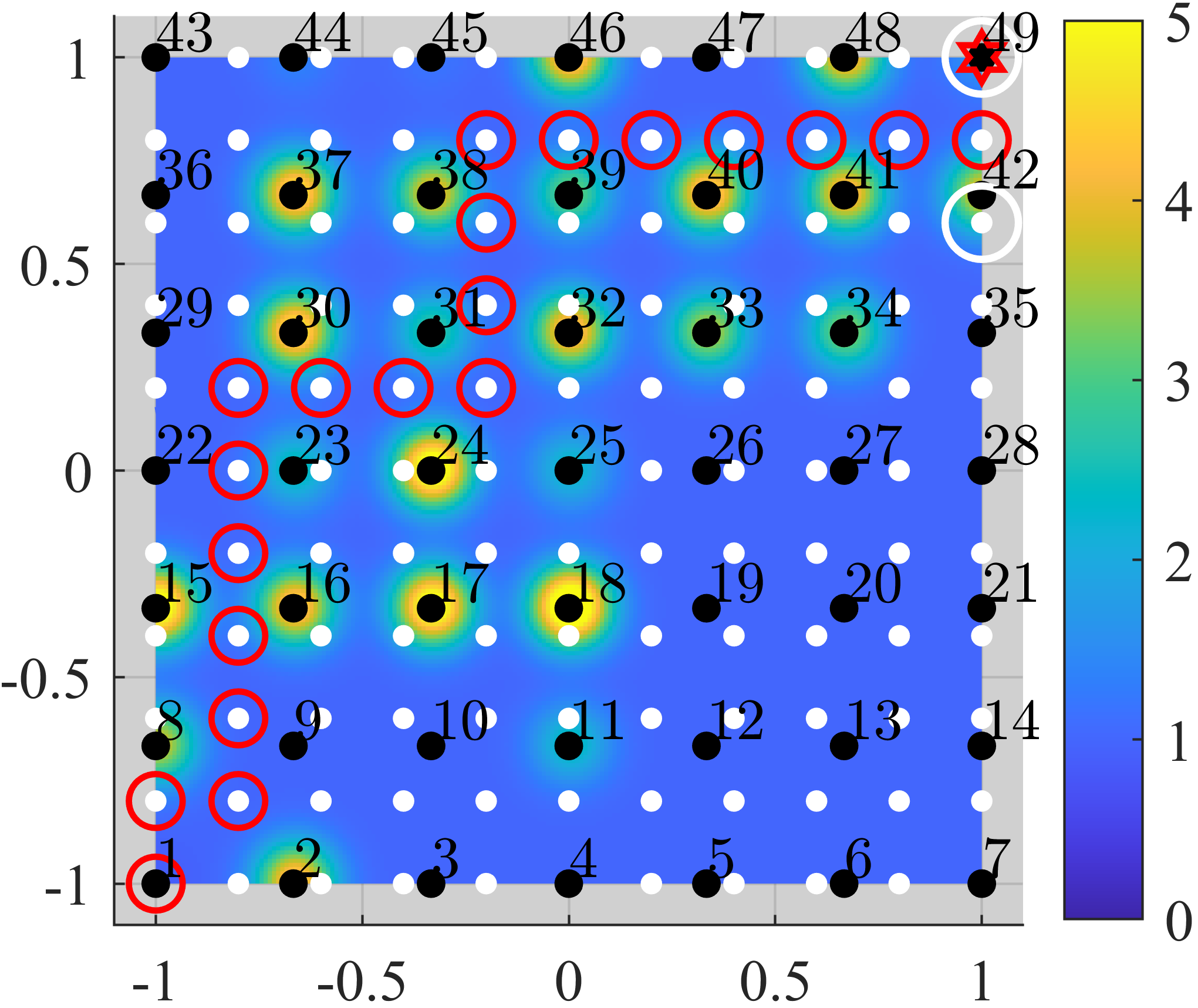}
	}
	\caption{Visualization of A-CSCP process for $N_{P} = 49$ and $\nGridPt = 121.$}
	\label{fig:Fig.2}
\end{figure}

An illustrative example of the A-CSCP algorithm with
$\constB=1$ is presented in \fig{fig:Fig.2}. Recall that with $\constB =1$ 
the sensor reconfiguration cost considers only the $d_1$ component.
The number of threat parameters, grid points,
and sensors are $N_{P}=49$, $\nGridPt=121$, and $\nSensor=2$, respectively.
The threat parameters $N_{P}$, indicated by the black dots and numbered 
from 1 to 49, are uniformly 
spaced in the workspace. The white dots represent the grid points, whereas
the white circles indicate sensor locations.
The start and goal locations are the bottom left and the top right grid points, respectively. 
The ego vehicle moves at a constant speed of 0.01 and both sensors move 
at a constant speed of 0.05, such that $\frac{\speedsensor}{\speedactor}=5$.

The evolution of the threat field estimate $\widehat{c}$ and the path 
progression of the ego vehicle at different reached vertices, 
namely $i=1, 6, 13$, and $19$ is shown in a color map,
where $i$ is the number of vertices traveled by the ego vehicle along 
the path.
The path traveled by the ego vehicle is represented by red circles, 
while the red stars indicate the planned future path. 
The planned path may be updated when new threat estimates become available. 
For $i=1$, a path is planned based on only the initial sensor measurements. 
As the vehicle reaches each vertex along the path, the next vertex 
is chosen based on the latest estimate of the threat field provided by the sensors. This progression is 
illustrated in \fig{fig:Fig.2} (b), (c) and (d).  

\def\thisfigwidth{0.46\columnwidth}
\begin{figure}
	\centering
	\subfigure[$\constA=0$]{
		\includegraphics[width=\thisfigwidth]{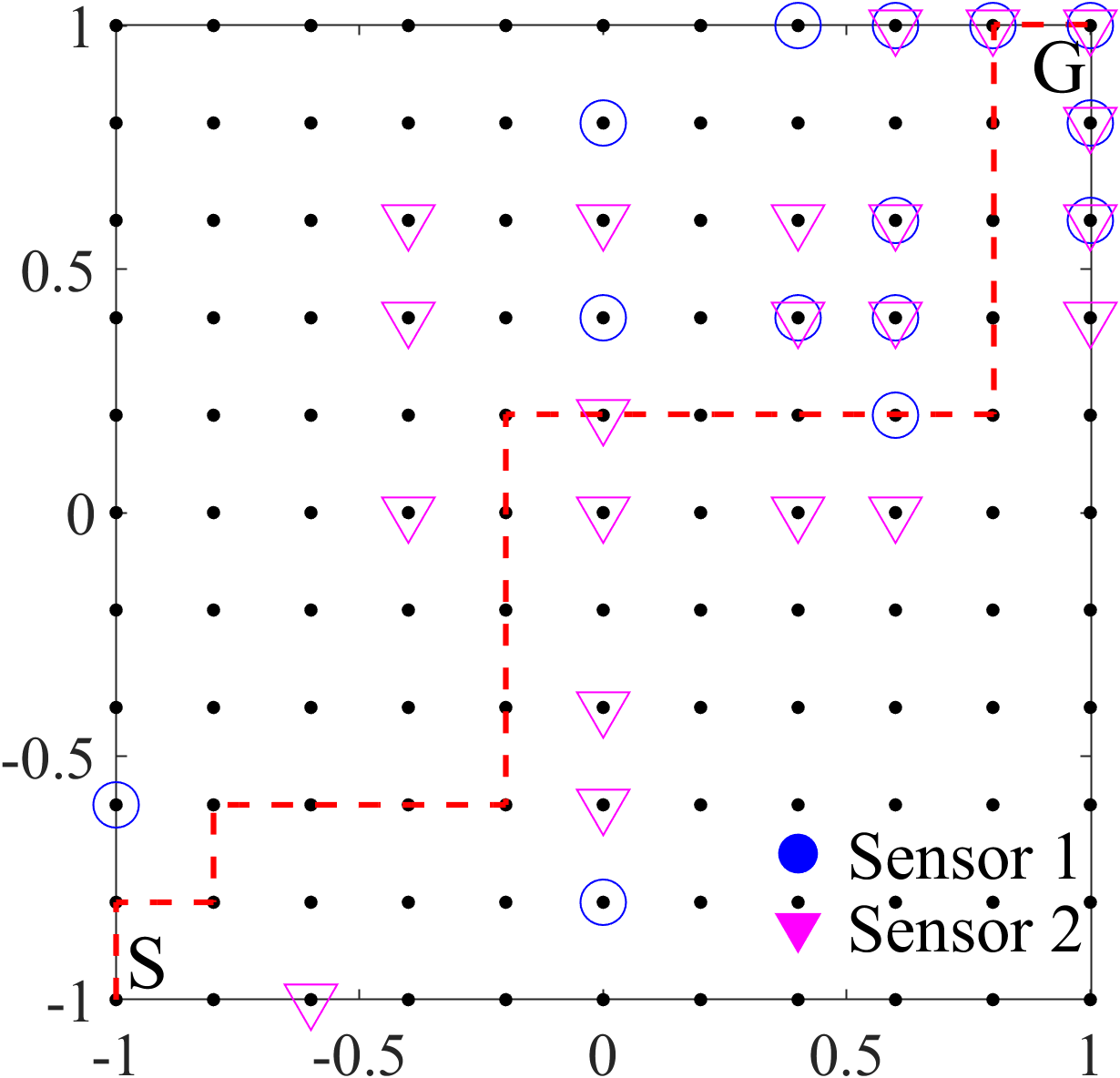}
	}
	\subfigure[$\constA\neq 0, \constB=1$]{
		\includegraphics[width=\thisfigwidth]{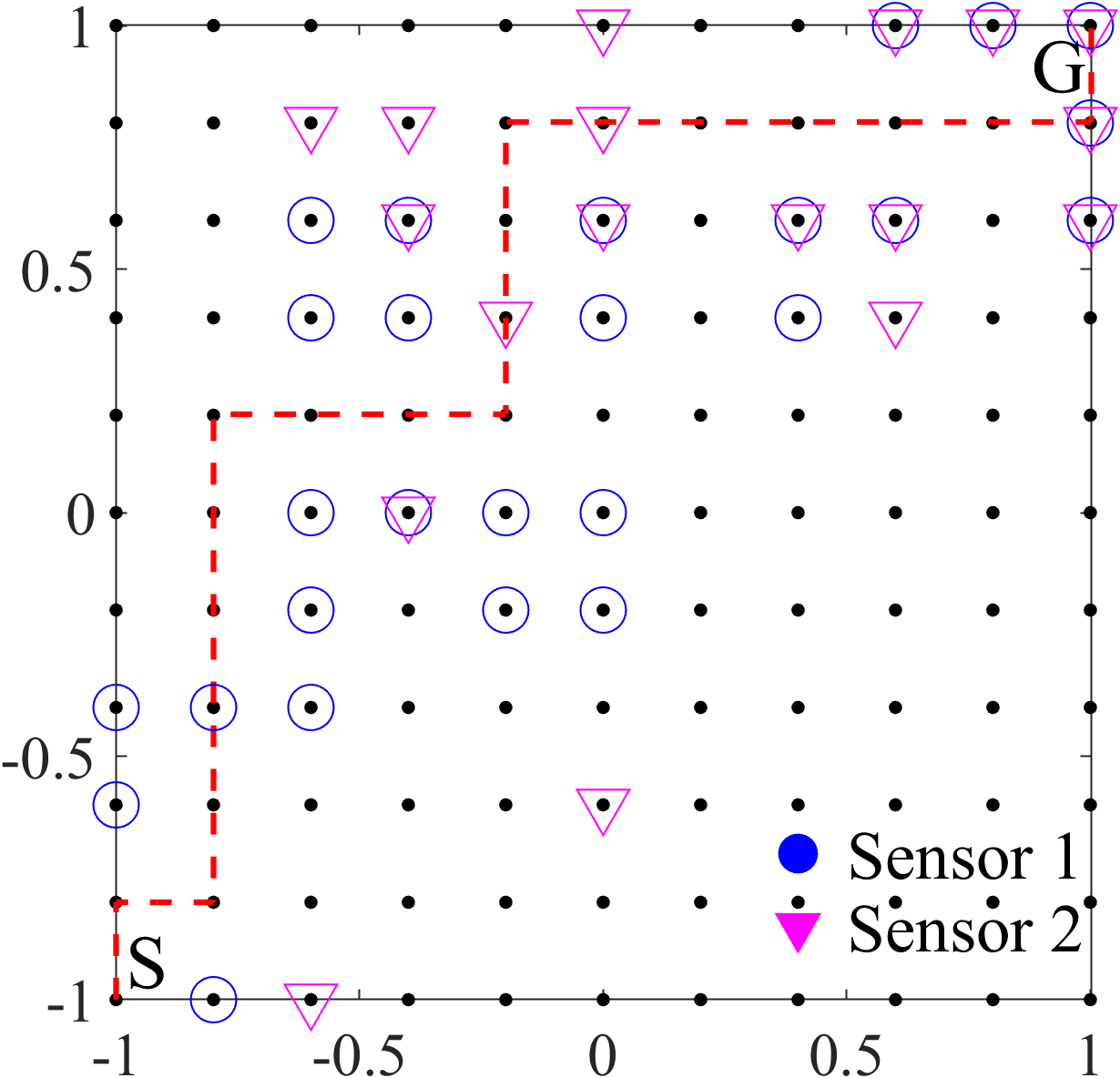}
	}
	\subfigure[$\constA\neq 0, \constB= 0.5$]{
		\includegraphics[width=\thisfigwidth]{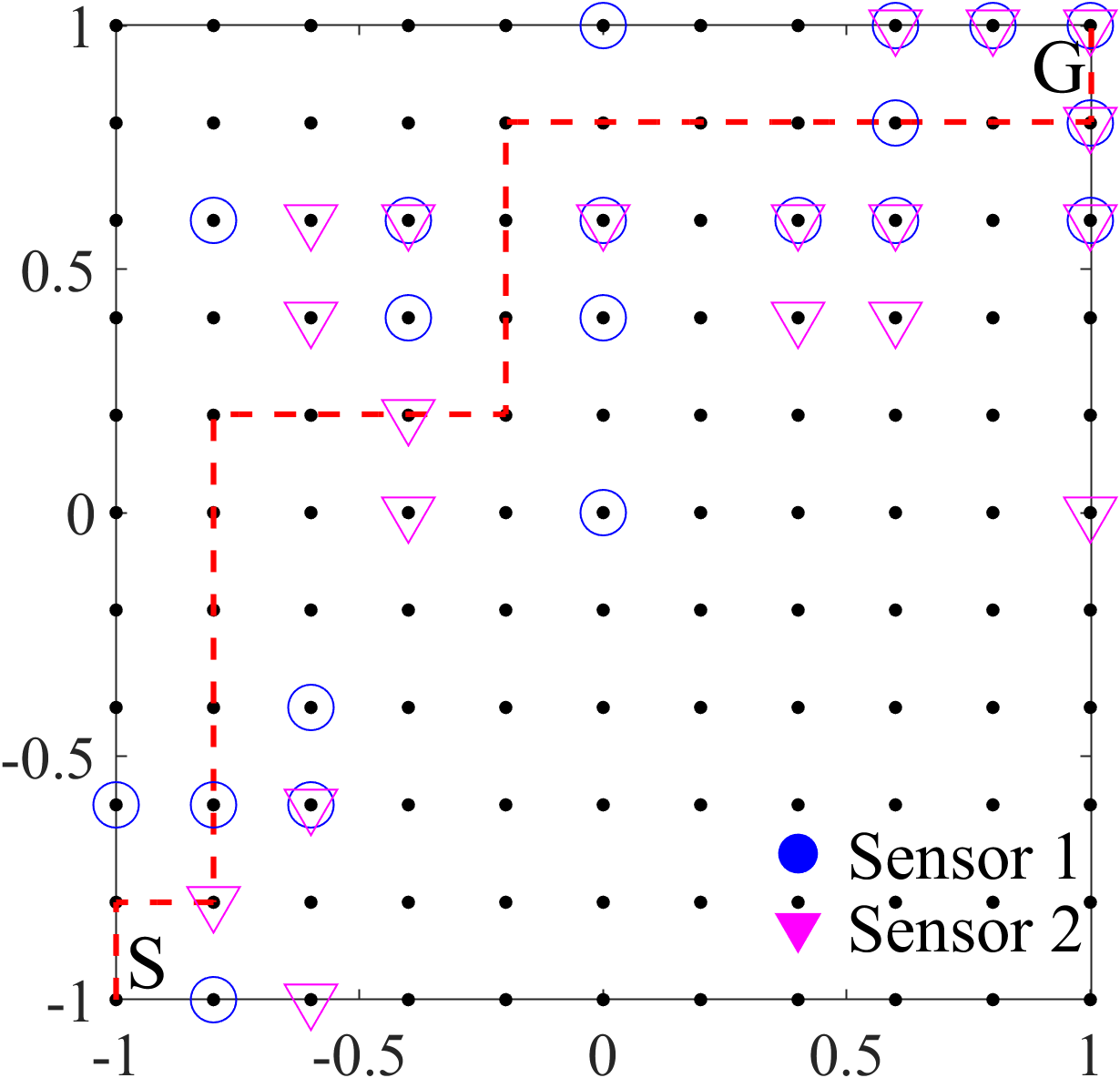}
	}
	\subfigure[$\constA\neq 0, \constB=0$]{
		\includegraphics[width=\thisfigwidth]{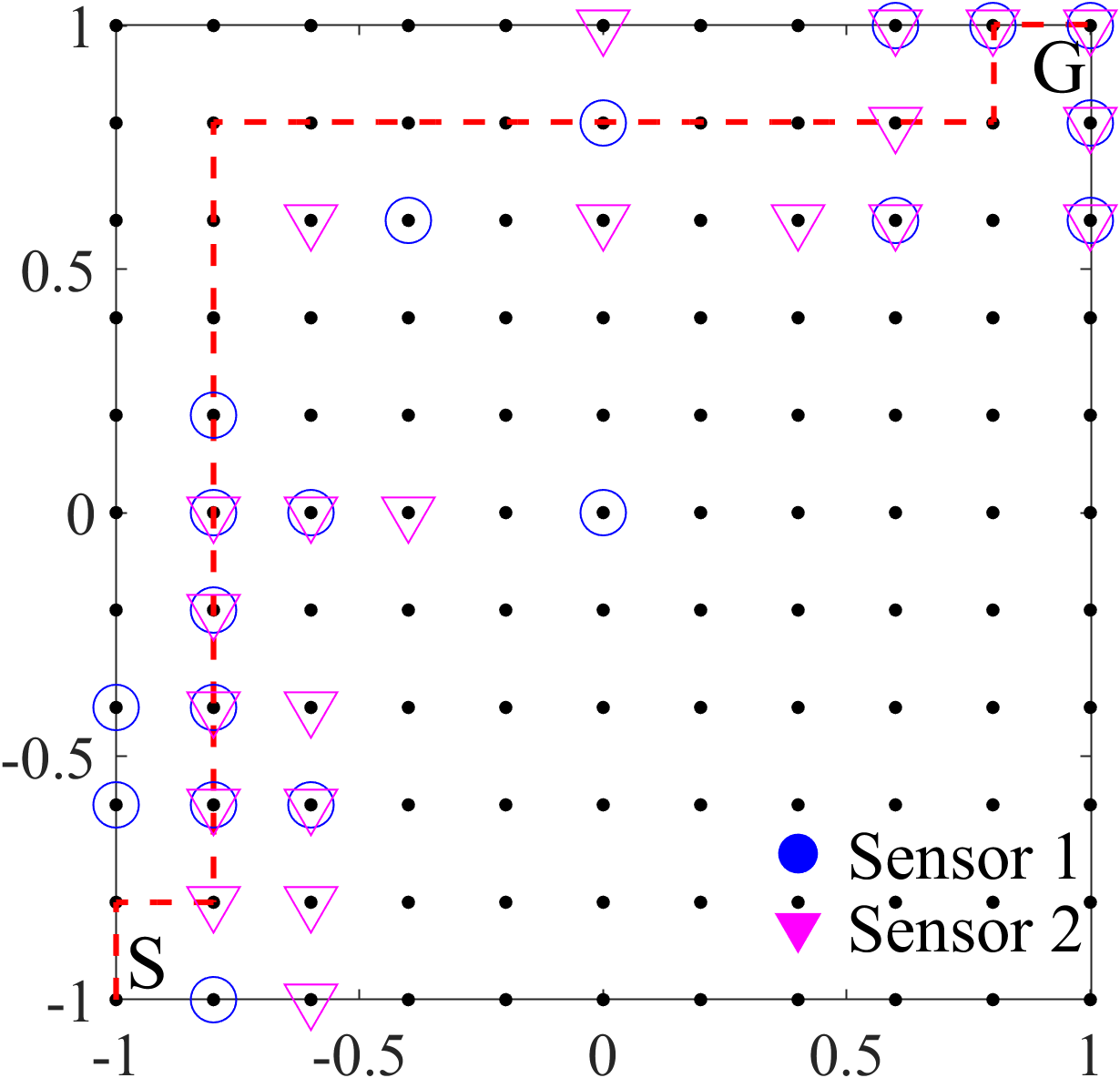}
	}
	\caption{Illustration of ego vehicle's path and sensor history for different sensor configuration schemes.}
	\label{fig:sensorpath}
\end{figure}

\subsection{Comparative Study}

For a comparative analysis, we implement via numerical simulation
the A-CSCP method on the same threat field as in \fig{fig:Fig.2} by 
varying the type of reward function $\SensorConfigReward$ used for
finding the optimal sensor configurations.

In these simulations we record not only the cost $J$ incurred by the
ego vehicle as it moves from one grid point to the next, but also the
\emph{total} exposure to the threat while traveling 
between the vertices.

Specifically, we record the \mydef{incurred cost}
\begin{align}
	\integralCost \defeq  \int_{\gridPath} 
	\boldsymbol\Phi^{\intercal}(\xState(t))\paramVec(t) \df t,
	\label{eq-totalthreatExposure}
\end{align}
where $\int_{\gridPath}$ is understood as the integral along a smooth curve
interpolated over the grid points in the path $\gridPath.$

For the sake of analyzing the quality of the path eventually followed
by the ego vehicle due to the proposed A-CSCP algorithm, consider two 
other benchmark paths for comparison: the worst-case path 
$\gridPath\msup{w}$ and the true optimal $\gridPath\msup{t}.$ 
Both of these benchmark paths are found assuming complete a priori knowledge
of the threat field (which the A-CSCP does not have). 
The worst-case path $\gridPath\msup{w}$ is chosen such that it passes 
through every peak of the threat field. The true optimal path $\gridPath\msup{t}$ 
is easily calculated using Dijkstra's algorithm with complete 
knowledge of the threat field. Note that, the path length for $\gridPath\msup{w}$ is same as of $\gridPath\msup{t}$. Let $\integralCost\msup{w}$ and $\integralCost\msup{t}$ denote the 
incurred costs of the worst-case and true optimal paths, respectively.

During the execution of the A-CSCP algorithm, we record the sensor
configurations and specifically we record the number of \emph{unique}
grid point locations at which sensors are placed. Let $S$ and $U$ denote
the total number and the unique number of grid points, respectively,
at which sensors are placed throughout the execution of A-CSCP.

An efficiency measure $\eta$ is defined as follows based on the ego vehicle's total 
threat exposure $\integralCost$ and the number of sensor placement locations:
\begin{align}
	\eta \defeq \left( 
	\frac{\integralCost\msup{w} - \integralCost}
	{\integralCost\msup{w} - \integralCost\msup{t}}\right)\frac{U}{S}.
	\label{eq-efficiencymetric}
\end{align}
This efficiency compares the incurred cost of the A-CSCP resultant path
to that of the worst-case path, normalized by the difference between the
worst-case and optimal incurred costs. $\eta$ also incorporates a measure
of efficiency in sensor motion, in that the efficiency is higher with fewer
repeated placements of sensors at the same grid points.

The first comparative reconfiguration scheme utilizes CRMI metric, 
and $\constA=0,$ which neglects any sensor placement costs. 
The other schemes consider the placement metrics that include 
$\SensorCost$, with different values of $\constB$, 
namely $\constB=1, 0.5$, and $0$.

\figf{fig:sensorpath} 
depicts the total path taken by the ego vehicle as well as the vertices 
visited by each sensor. For $\constA=0$,  
the sensors have a tendency to travel farther between successive configurations, 
thereby taking fewer 
measurements by the time ego vehicle reaches the goal. 
This can be seen when comparing the $S$ values 
from \tbl{tab:sensor_locations_visted}. 
Furthermore, some grid locations have inherently higher MI values 
such as the goal, so there is a tenancy to frequently visit these locations. 
The result of these two factors is 
evident in \fig{fig:sensorpath}(a). The effect of the different placement schemes 
may be visually compared visually. For example, \figf{fig:sensorpath}(d) 
shows how the sensors strictly adhere to the path and rarely deviate, where 
\figf{fig:sensorpath}(b) shows a much larger portion of the threat field being explored.

Table \ref{tab:threat_exposure} provides the ego vehicle's normalized threat exposure.
To compute the normalized values, the benchmark incurred costs
$\integralCost\msup{t}$ and $\integralCost\msup{w}$ are assigned normalized
scores of 1 and 0, respectively. That is, normalized values close to 1 indicate 
near-optimality.
The schemes are tested for different sensor to ego vehicle speed ratios. 
The sensor speed will inherently impact the threat field exploration. 
For example, if the sensor speed is much greater than the ego vehicle speed, 
there is ample time for more of the threat field to be explored, and the sensor 
reconfiguration scheme becomes trivial. Table~\ref{tab:sensor_locations_visted} 
shows for $\frac{\speedsensor}{\speedactor} = 5$ the values of
$S$ and $U$ recorded regarding sensor efficiency and the efficiency measure 
$\eta$ as defined in~\eqnnt{eq-efficiencymetric}.
A scheme that finds the optimal path without the sensors revisiting 
any vertices achieves $\eta = 1$. The inclusion of $\SensorCost$ always 
increases $\eta$, with $\constB = 1$ performing the best.

\renewcommand{\arraystretch}{1.5}

\begin{table}[h!]
	\centering
	\caption{Normalized Threat Exposure ($\integralCost$)}
	\begin{tabular}{|c|p{1.7cm}|p{1.4cm}|p{1.4cm}|p{1.4cm}|} 
		\hline
		\multirow{2}{*}{$\frac{\speedsensor}{\speedactor}$} & \multirow{2}{*}{\textbf{CRMI} ($\constA=0$)} & 
		\multicolumn{3}{c|}{\textbf{CRMI + Cost} ($\constA\neq 0$)} \\ \cline{3-5} 
		&  & \textbf{$\constB=1$} & \textbf{$\constB=0.5$} & \textbf{$\constB=0$} \\ \hline
		5 & 0.8712 & 0.9948 & 0.8974 & 0.9384 \\ \hline
		10  & 0.9064 & 0.9948 & 0.9841 & 0.9917 \\ \hline
		50  & 1 & 1 & 0.9948 & 1 \\ \hline
	\end{tabular}
	\label{tab:threat_exposure}
\end{table}

\begin{table}[h!]
	\centering
	\caption{Efficiency Metric ($\eta$)}
	\begin{tabular}{|c|p{1.7cm}|p{1.3cm}|p{1.3cm}|p{1.3cm}|} 
		\hline
		\multirow{2}{*}{} & \multirow{2}{*}{\textbf{CRMI} ($\constA=0$)} & \multicolumn{3}{c|}{\textbf{CRMI + 
		Cost} ($\constA\neq 0$)} \\ \cline{3-5} 
		&  & \textbf{$\constB=1$} & \textbf{$\constB=0.5$} & \textbf{$\constB=0$} \\ \hline
		 $S$ & 69 & 89 & 75 & 83 \\ \hline
		$U$ & 27 & 34 & 29 & 29 \\ \hline
		$\eta$ & 0.3409 & 0.3800 & 0.3470 & 0.3278 \\ \hline
	\end{tabular}
	\label{tab:sensor_locations_visted}
\end{table}

\section{Conclusions}
\label{sec-conclusions}

In this paper, we presented a technique for simultaneously sensing and planning
for an ego vehicle moving at a constant speed, accompanied by multiple sensors also 
moving at constant speeds. The sensor placement is based on maximizing a reward 
that includes uncertainty reduction in the path cost and reduction in sensor movement. 
A sensor reconfiguration cost is considered for reducing the distance traveled by 
sensors between configurations while ensuring that the sensors remain in proximity
to the ego vehicle. Numerical simulations are performed on different schemes of 
sensor placement metric and various sensor to ego vehicle speed ratios. 
The performance efficiency of the different schemes is evaluated based on the 
total threat exposure and the number of sensor configurations during the A-CSCP 
process. Based on the defined performance metric, the sensor placement scheme that 
maximizes the CRMI metric while minimizing the sensor travel distance between 
successive configurations performs better in terms of an efficiency measure.

\addtolength{\textheight}{-20\baselineskip}

\section*{Acknowledgments}
This work is funded in part by NSF grant \#2126818.

\bibliographystyle{ieeetr}
\bibliography{References}

\end{document}